\documentclass{article}

\title{\bf Are randomness of behavior and information flow important to opinion forming in organization?}
\author{Agnieszka Kowalska-Styczeń$^1$\thanks{Agnieszka.Kowalska-Styczen@polsl.pl}, Krzysztof Malarz$^2$\thanks{malarz@agh.edu.pl}}
\date{{\small
$^1$Silesian University of Technology,
Faculty of Organisation and Management\\
ul. Roosevelta 26/28, 41-800 Zabrze, Poland\\[1ex]
$^2$AGH University of Science and Technology,
Faculty of Physics and Applied Computer Science\\
al. Mickiewicza 30, 30-059 Krak\'ow, Poland}\\[2ex]
\today
}

\usepackage{amsmath}
\usepackage{graphicx}
\usepackage{subcaption}
\usepackage[round]{natbib}
\usepackage[left=3cm,right=3cm,top=3cm,bottom=3cm]{geometry}
\usepackage[colorlinks=true]{hyperref}
\usepackage{cleveref}

\begin{document}

\maketitle

\begin{abstract}
    We examine how the randomness of behavior and the flow of information between agents affect the formation of opinions. Our main research involves the process of opinion evolution, opinion clusters formation and studying the probability of sustaining opinion. The results show that opinion formation (clustering of opinion) is influenced by both flow of information between agents (interactions outside the closest neighbors) and randomness in adopting opinions.
\end{abstract}

\section{Introduction}

Understanding what factors influence the formation of opinion in society is very important for many aspects of an organization's activities, including organizational behavior, organizational knowledge transfer, leadership and many more. The basis of the proposed approach in this paper is to look at the organization as a complex social system. This direction of research is more and more often seen as an important element of management science \citep{Kowalska2018}. To explain the specifications of social systems (including organization) such as self-organization, order, chaos, complexity, the evolution of mathematical theories must be extended \citep{Ehsani2010}. Such an extension of the theory is the usage of computer modeling, simulation, in particular agent-based modeling \citep{Kowalska2020}.
As emphasized by \cite{Acemoglu-2011}, we acquire our beliefs and opinions through various types of experiences, in particular through the process of `social learning'. As part of this process, people communicate with other people, they obtain information and they update their beliefs and opinions. The communication takes place with a subgroup of society consisting of friends, colleagues and peers, co-workers and distant and close family members. Information obtained by a person from a particular partner on a social network is transferred to other members of that network. Depending on the communication (information flow) in the social network, this information transfer may have a smaller or larger reach, the information may amount to more or less people. It should also be noted that in the formation of opinions, we often deal with unpredictability, as well as with irrational processing of information \citep{Sobkowicz-2018,Stadelmann2013}. 

The purpose of this article is to study the formation of clusters of opinions (i.e. groups of people with the same opinion). In a social network, individual opinions and interpersonal relationships always interact and evolve, leading to self-organization of clusters of opinions across the network \citep{Zhang2013}. Because it is very difficult to study social systems in terms of their complexity, researchers increasingly use computer simulations. Therefore, in this article, we propose simulations carried out  for a model based on \cite{Latane-1981} theory of social impact, which takes into account both the different flow of information between community members and the randomness of their behavior (noise). The basic assumption of Latan\'e's social impact theory is that people are members of a community and within that community they interact with each other. Social influence are all interactions between people (persuasion, joke, showing emotions, feelings, etc.). The theory of social influence is based on three principles: social power, psychosocial law, and multiplication or division.

\section{Model description}

The computer model used here has been formally described in the work by \cite{1902.03454}. The model was also tested by \cite{2002.05451} for two and three  opinions available in the system and initially randomly distributed among agents. The formation of opinions in the community as a result of the flow of information and randomness of agents behavior was investigated, and several phases in the behavior of the system (community) due to these parameters were detected. 
In this paper, we focused on the relationship between the spatial distribution of opinions and the  probability of sustaining opinion by agents in opinion clusters. 

It should be noted that, the essence of social influence is  not only exerting social influence, but also succumbing to it. In the used model, this fact was taken into account by introducing the following parameters: intensity of persuasion and intensity of support. These parameters determine the effectiveness of which an individual may interact with or influence other individuals by changing or confirming their opinions. These parameters are interpreted as follows:
\begin{itemize}
\item intensity of persuasion: the larger $0\le p_i\le 1$ the agent is more convincing agents with other opinions from his neighborhood to accept his/her opinion,
\item intensity of support: the larger $0\le  s_i\le 1$ the agent convinces more strongly with other agents from his neighborhood, so that they do not change their opinion if this opinion is identical with the opinion of agent $i$.
\end{itemize}

In the simulations presented in the next section, random values of $p_i$ and $s_i$ for each agent were assumed.
As in previous works, we take into account two parameters of the model: information flow in the community and random behavior (noise). The varied flow of information in the community was expressed in the model by the parameter $\alpha$ (this parameter talks about the influence of close or distant neighbors in the community). Small values (e.g. $\alpha=1$ or 2) mean good communication between agents and good access to information (an exchange of information with a large number of agents in the network takes place). The larger values of $\alpha$, the weaker the communication between the groups of agents (the exchange of information takes place only in the closest neighborhood, e.g. $\alpha = 6$).

The randomness of agents' behavior is expressed by the parameter $T$ (social temperature). If $T=0$, then a lack of randomness is observed, and the agent adopts an opinion that has the most influence on him/her. As $T$ increases, more and more often opinions that do not have the greatest impact on agents are chosen.

To sum up, in the used computer model:
\begin{itemize}
\item	The society is represented by agents occupying the nodes of the square lattice. Each network node corresponds to one person.
\item	Every agent $i$ has one among $K$ available opinions and, moreover, it is characterized by two parameters: intensity of persuasion ($p_i$) and intensity of support ($s_i$). We assume random values of these parameters for all agents.
\item	Each agent is influenced by all other agents. The strength of this influence decreases with increasing distance between them. Additionally, in order to reduce the impact of distant neighbors, the parameter $\alpha$ was introduced, which appears as a constant in the distance scaling function.
\item	The randomness of human behavior is reflected by  $T$ parameter, which   characterize the entire system.
\item The simulations are carried out on square lattice of linear size $L=41$ with open boundary conditions.
\end{itemize}

\section{Results}

\subsection{Spatial distribution of opinions} 
\begin{figure*}[!htbp]
\centering
\begin{subfigure}[b]{.32\textwidth} \caption{\label{fig:xi-a} $\alpha=2$, $T=0$}
	\includegraphics[width=45mm]{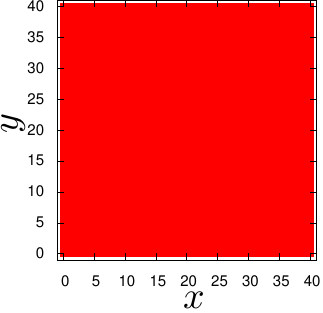}
\end{subfigure}
\begin{subfigure}[b]{.32\textwidth} \caption{\label{fig:xi-b} $\alpha=3$, $T=0$}
	\includegraphics[width=45mm]{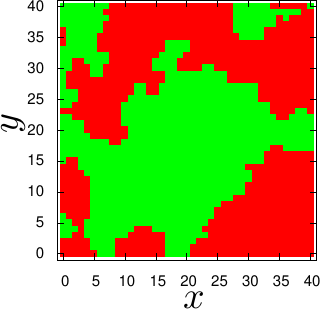}
\end{subfigure}
\begin{subfigure}[b]{.32\textwidth} \caption{\label{fig:xi-c} $\alpha=6$, $T=0$}
	\includegraphics[width=45mm]{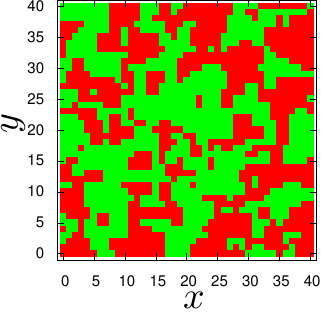}
\end{subfigure}
\begin{subfigure}[b]{.32\textwidth} \caption{\label{fig:xi-d} $\alpha=2$, $T=1$}
	\includegraphics[width=45mm]{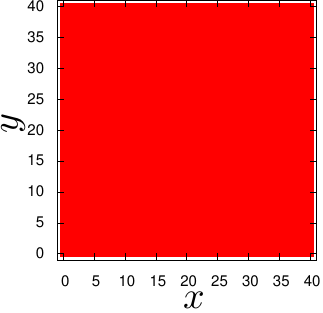}
\end{subfigure}
\begin{subfigure}[b]{.32\textwidth} \caption{\label{fig:xi-e} $\alpha=3$, $T=1$}
	\includegraphics[width=45mm]{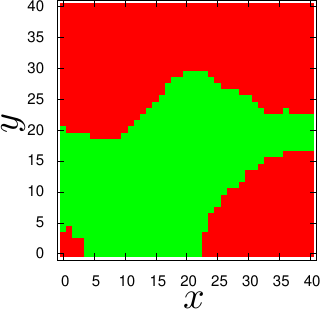}
\end{subfigure}
\begin{subfigure}[b]{.32\textwidth} \caption{\label{fig:xi-f} $\alpha=6$, $T=1$}
	\includegraphics[width=45mm]{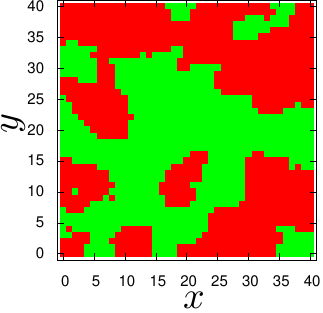}
\end{subfigure}
\begin{subfigure}[b]{.32\textwidth} \caption{\label{fig:xi-g} $\alpha=2$, $T=2$}
	\includegraphics[width=45mm]{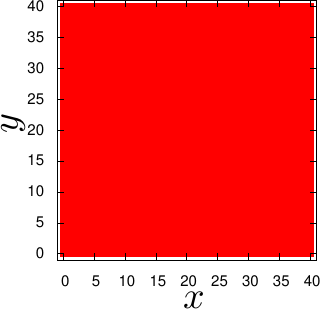}
\end{subfigure}
\begin{subfigure}[b]{.32\textwidth} \caption{\label{fig:xi-h} $\alpha=3$, $T=2$}
	\includegraphics[width=45mm]{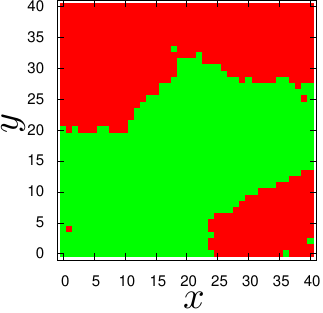}
\end{subfigure}
\begin{subfigure}[b]{.32\textwidth} \caption{\label{fig:xi-i} $\alpha=6$, $T=2$}
	\includegraphics[width=45mm]{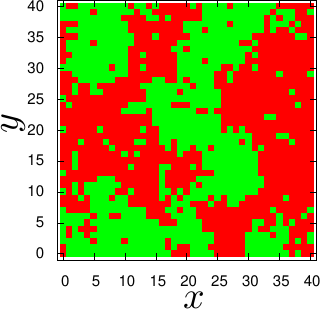}
\end{subfigure}
\begin{subfigure}[b]{.32\textwidth} \caption{\label{fig:xi-j} $\alpha=2$, $T=3$}
	\includegraphics[width=45mm]{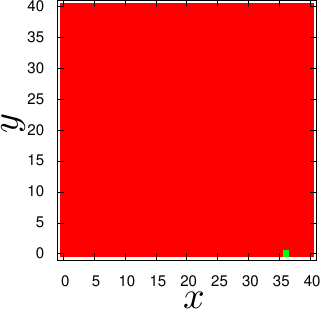}
\end{subfigure}
\begin{subfigure}[b]{.32\textwidth} \caption{\label{fig:xi-k} $\alpha=3$, $T=3$}
	\includegraphics[width=45mm]{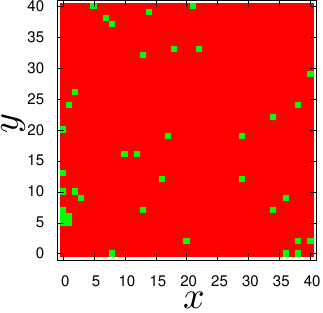}
\end{subfigure}
\begin{subfigure}[b]{.32\textwidth} \caption{\label{fig:xi-l} $\alpha=6$, $T=3$}
	\includegraphics[width=45mm]{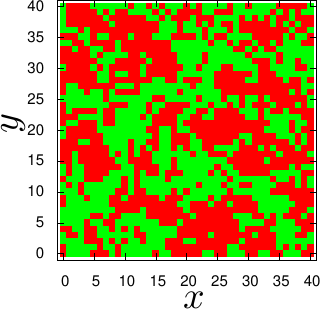}
\end{subfigure}
\caption{\label{fig:xi}Spatial distribution of opinions for $K=2$ opinions for various $\alpha$ and various social temperatures ($T$) after $t=100$ time steps of the system evolution (red and green mean different opinions). $L=41$.}
\end{figure*}

Let us start presentation of the results by showing the spatial distribution of opinions for $K=2$ (two possible opinion), for $\alpha=2$, 3, 6 (various levels of flow of information) and for $T=0$, 1, 3, 5 (various levels of noise, i.e. various levels of randomness in the opinion adopting).
As can be seen in \Cref{fig:xi}, both $\alpha$ and social temperature $T$ have influence on opinion formation and polarization in the groups (self-organization of opinion clusters). 

For $\alpha = 2$ (and $\alpha = 1$) the consensus takes place (all agents accept one of two opinions), except of single cases for large $T$ values. The higher $\alpha$, the less interactions unit has outside the closest neighbors group, which leads to lower polarization of opinions (there are more and more clusters, i.e. groups of agents with the same opinions). In addition, small clusters are able to survive, although they are surrounded by large clusters of agents with different opinion. An analogous situation occurs when the $T$ parameter is increased. With larger $T$, there is more heterogeneity in the areas where actors with opposite opinions occur and the division into clusters is less pronounced (less polarization of opinions in the groups). An increase of the social temperature $T$ often results in the emergence of small clusters of agents with a minority of opinion. In general, the increase of $T$ and  causes that more clusters are formed, and the polarization of opinions in groups is weaker.

\subsection{Probability of sustaining opinion} 
\begin{figure*}[!htbp]
\centering
\begin{subfigure}[b]{.32\textwidth} \caption{\label{fig:pso-a} $\alpha=2$, $T=1$}
	\includegraphics[width=45mm]{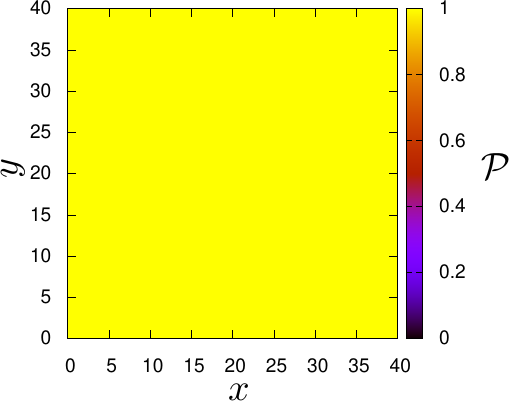}
\end{subfigure}
\begin{subfigure}[b]{.32\textwidth} \caption{\label{fig:pso-b} $\alpha=3$, $T=1$}
	\includegraphics[width=45mm]{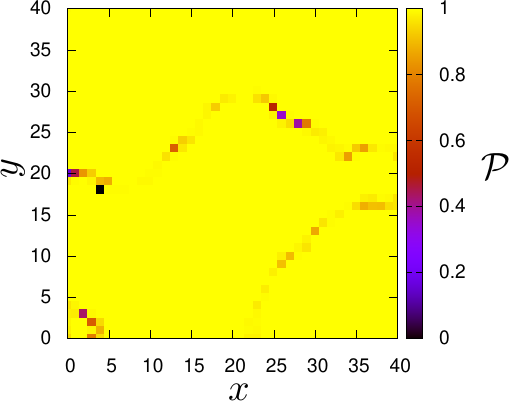}
\end{subfigure}
\begin{subfigure}[b]{.32\textwidth} \caption{\label{fig:pso-c} $\alpha=6$, $T=1$}
	\includegraphics[width=45mm]{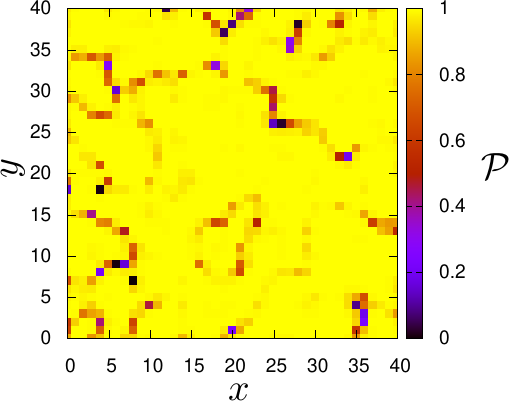}
\end{subfigure}	
\begin{subfigure}[b]{.32\textwidth} \caption{\label{fig:pso-d} $\alpha=2$, $T=2$}
	\includegraphics[width=45mm]{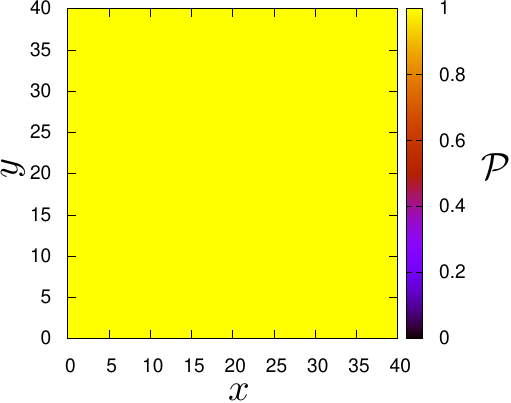}
\end{subfigure}
\begin{subfigure}[b]{.32\textwidth} \caption{\label{fig:pso-e} $\alpha=3$, $T=2$}
	\includegraphics[width=45mm]{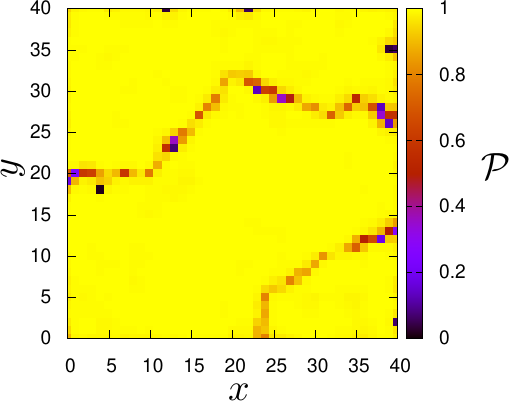}
\end{subfigure}
\begin{subfigure}[b]{.32\textwidth} \caption{\label{fig:pso-f} $\alpha=6$, $T=2$}
	\includegraphics[width=45mm]{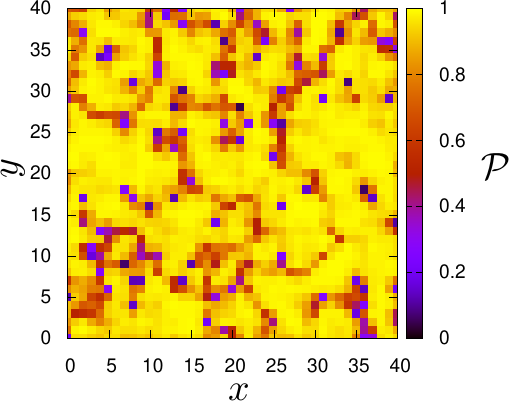}
\end{subfigure}
\begin{subfigure}[b]{.32\textwidth} \caption{\label{fig:pso-g} $\alpha=2$, $T=3$}
	\includegraphics[width=45mm]{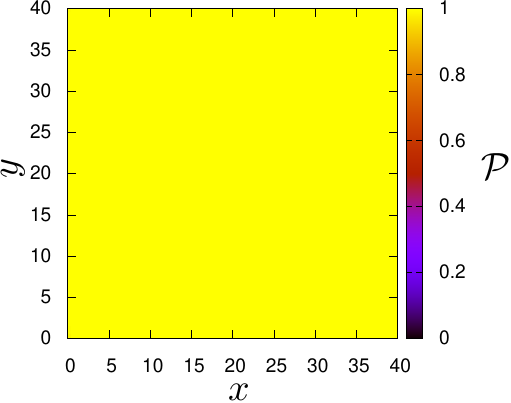}
\end{subfigure}
\begin{subfigure}[b]{.32\textwidth} \caption{\label{fig:pso-h} $\alpha=3$, $T=3$}
	\includegraphics[width=45mm]{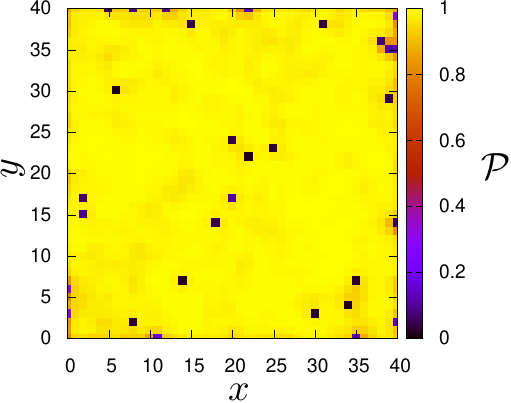}
\end{subfigure}
\begin{subfigure}[b]{.32\textwidth} \caption{\label{fig:pso-i} $\alpha=6$, $T=3$}
	\includegraphics[width=45mm]{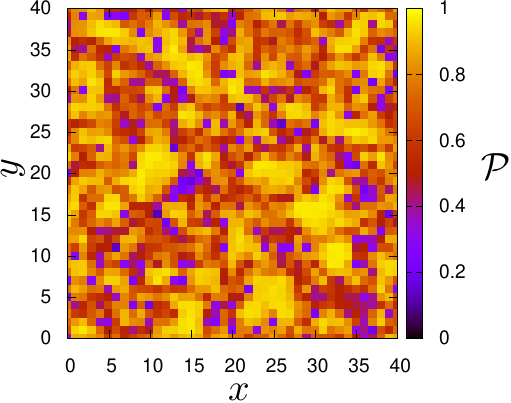}
\end{subfigure}
\begin{subfigure}[b]{.32\textwidth} \caption{\label{fig:pso-j} $\alpha=2$, $T=5$}
	\includegraphics[width=45mm]{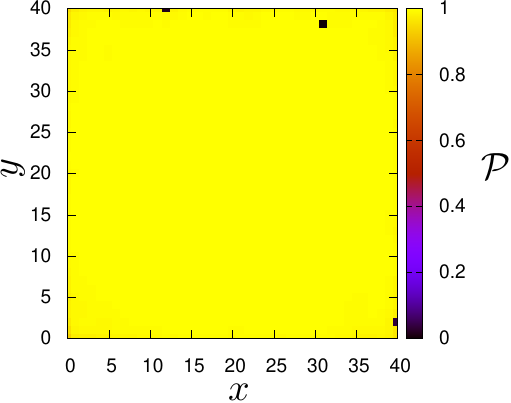}
\end{subfigure}
\begin{subfigure}[b]{.32\textwidth} \caption{\label{fig:pso-k} $\alpha=3$, $T=5$}
	\includegraphics[width=45mm]{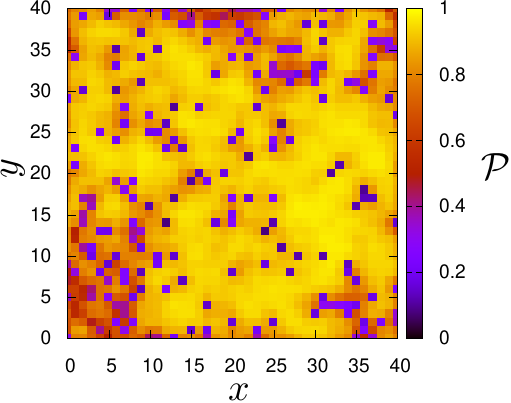}
\end{subfigure}
\begin{subfigure}[b]{.32\textwidth} \caption{\label{fig:pso-l} $\alpha=6$, $T=5$}
	\includegraphics[width=45mm]{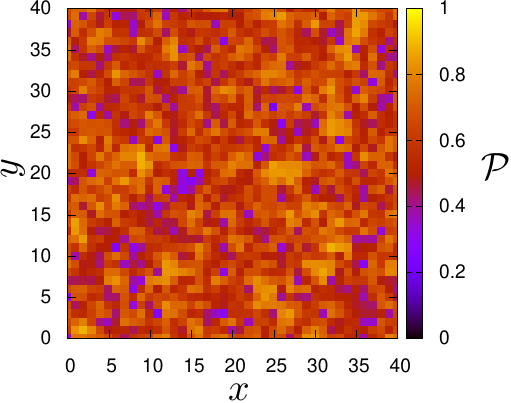}
\end{subfigure}
	\caption{\label{fig:pso}Spatial distribution of probabilities of sustaining opinion  for $K=2$ opinions, social temperature $T$ and for various values of $\alpha$ after $t=100$ time steps of the system evolution. $L=41$.}
\end{figure*}

For \Cref{fig:xi} discussed in the previous Section, corresponding heat-maps have been created (see \Cref{fig:pso}). Each agent is assigned to one point in the network with a certain color. This color depends on the probability that agent will sustain his/her opinion. The colors change from yellow (high probability of sustaining opinion) to black (low probability of sustaining opinion). We do not show heat-maps for $T=0$, because the probability of sustaining the opinion has then only two values: either 0---the agent will change his/her opinions or 1---the agent will sustain his/her opinion. The smallest probability of sustaining the opinion occurs for agents on the outskirts of clusters---compare \Cref{fig:pso} to \Cref{fig:xi}. In addition, this probability is higher in large clusters than in small ones. This likelihood also decreases with the increase of $T$ (more and more darker colors in the agents’ network). This probability is of course affected by $T$ and  $\alpha$. The higher $T$, the less probability of sustaining the opinion (less yellow), which is especially visible for  $\alpha = 6$ in the \Cref{fig:pso-i} and \Cref{fig:pso-l}. The chances of sustaining the current opinion of agents also decrease with the rise of $\alpha$ (there is more and more darker color). To sum up, the greater the randomness of behavior and the smaller the flow of information (i.e. greater $\alpha$), the less yellow color in the heat-maps is observed (a lower probability of sustaining opinion in the entire lattice of agents), which is particularly evident in \Cref{fig:pso-i} ($T=3$) and \Cref{fig:pso-k,fig:pso-l} ($T=5$).

\subsection{Clustering of opinions} 
Analyzing the results in the previous Section, the clustering of opinions is influenced by both the level of randomness in agents’ decisions ($T$) and the influence coming from close or distant neighbors ($\alpha$). For clusters detection, the \cite{Hoshen1976a} algorithm has been applied. In Hoshen--Kopelman algorithm each agent is labelled in such way, that agent with the same opinions and in the same cluster have identical labels. The algorithm in this paper was used for a square lattice with von Neumann neighborhoods (with the nearest neighbours interactions).

In \Cref{tab:Smax}, the relative size $\mathcal{S}_{\text{max}}/L^2$ of the largest cluster after $t=1000$ simulations time steps for different values of $T$ and  for $K = 2$ has been shown. These results coincide with \Cref{fig:xi}. In all cases, for $\alpha = 1$, the largest cluster fills the entire lattice in 100\% (i.e. there is a consensus in opinion). It should be noted that this is a situation in which most agents in the lattice influence the opinion of the selected agent (the flow of information between agents is very good). For $\alpha = 2$, the influence of other agents in the lattice is still greater than the influence of the closest neighborhood. The size of the maximum cluster then fluctuates around 100\% of the network size (i.e. one large cluster with one opinion). In general, for any size of $T$, the size of the largest cluster of opinions decreases with the increase of $\alpha$.

\begin{table}
\caption{\label{tab:Smax}The relative size of the largest cluster $\mathcal{S}_{\text{max}}/L^2$ for $L=41$.}
\begin{center}
\begin{tabular}{rrrrr}
\hline\hline
      & $\alpha=1$ &     2 &     3 &     6 \\ \hline
     $T=0$ & 100\% &  94\% &  54\% &  34\% \\
    	1  & 100\% & 100\% &  83\% &  57\% \\
	    2  & 100\% &  99\% &  92\% &  62\% \\
	    3  & 100\% &  99\% &  92\% &  24\% \\ 
	    4  & 100\% &  99\% &  90\% &  15\% \\  
	    5  & 100\% & 100\% &  85\% &  12\% \\ \hline\hline
\end{tabular}
\end{center}
\end{table}

\begin{figure}[!htbp]
\begin{center}
\begin{subfigure}[b]{.44\textwidth} \caption{\label{fig:his_clu_siz_21} $\alpha=1$}
	\includegraphics[width=60mm]{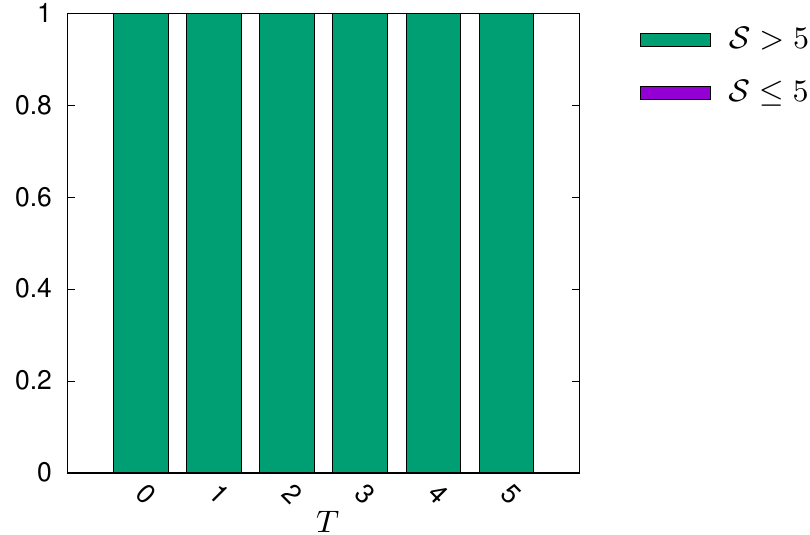}
\end{subfigure}
\begin{subfigure}[b]{.44\textwidth} \caption{\label{fig:his_clu_siz_22} $\alpha=2$}
	\includegraphics[width=60mm]{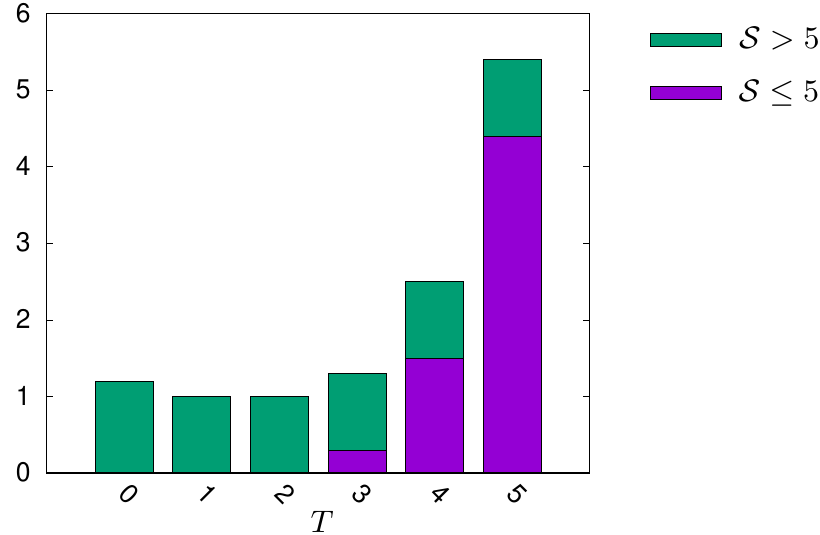}
\end{subfigure}\\
\begin{subfigure}[b]{.44\textwidth} \caption{\label{fig:his_clu_siz_23} $\alpha=3$}
	\includegraphics[width=60mm]{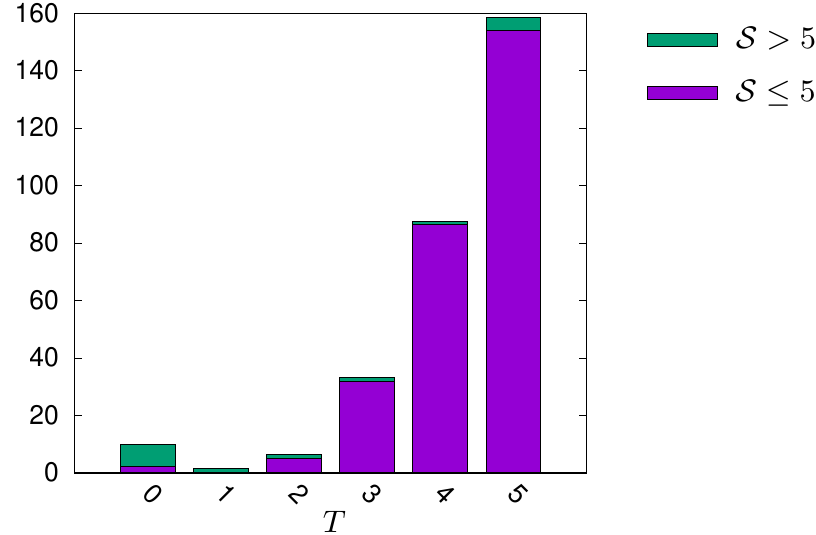}
\end{subfigure}
\begin{subfigure}[b]{.44\textwidth} \caption{\label{fig:his_clu_siz_26} $\alpha=6$}
	\includegraphics[width=60mm]{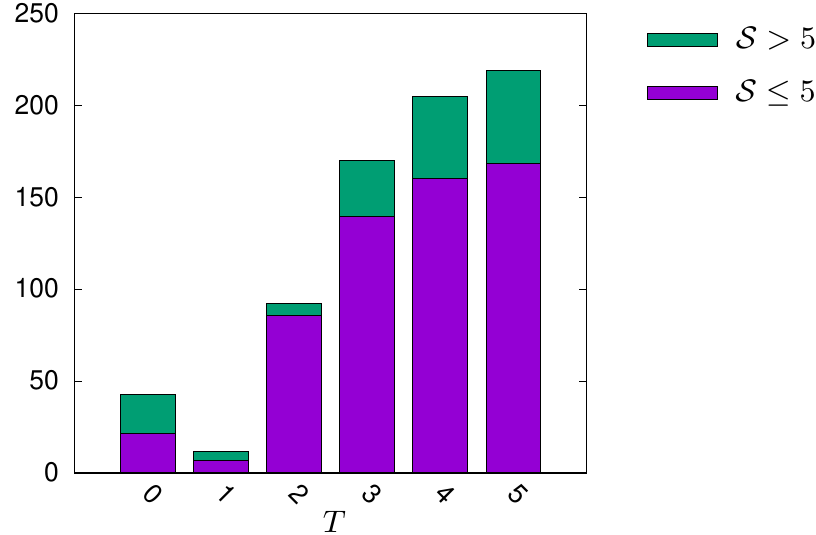}
\end{subfigure}
\end{center}
\caption{\label{fig:his_clu_siz}Histogram of cluster sizes $\mathcal{S}$ for various values of the temperature $T$ and parameter $\alpha$. $L=41$.}
\end{figure}

In order to get a better look at the clustering of opinions,  the average number of small and large clusters after thousand time steps for hundred simulations was also studied (see \Cref{fig:his_clu_siz}). We assumed that small clusters contain no more than five agents sharing the same opinion ($\mathcal{S} \le 5$). The randomness in accepting opinions ($T$) often results in the formation of small clusters of opposing opinions (as can be seen in the \Cref{fig:xi}). The number of clusters  increases with an increase of $\alpha$, i.e. the smaller the influence of all agents in the network on the selected agent, the more difficult is appearing the opinions clustering. In general, the number of clusters increases with $T$, but for $\alpha=3$ and $\alpha=6$ and for $T=1$, the number of clusters is lower than for $T = 0$. First of all, the probability of a new cluster forming (with opposite opinion) inside another cluster is still very small. Secondly, the probability of changing opinions is greater at the clusters borders  than inside the cluster. If the adjacent clusters have different sizes, then the probability of changing the agent’s opinion in a smaller cluster is greater than in the larger one. This situation leads to the disappearance of small clusters, or in general to a reduction in number of clusters. We can see similar phenomena in terms of the number of small clusters (see \Cref{fig:his_clu_siz}). The number of small clusters increases with $T$ (of course, apart from low $T$ values and for $\alpha>1$).

\section{Summary and conclusions}

In this article, we used a cellular automaton model in which the theories of Latan\'e's social influence were implemented. We studied how the social temperature (randomness of human behavior) and the flow of information in society influenced the clustering of opinion. To achieve this goal, the spatial distribution of opinions, probability of sustaining opinion and the number of clusters were studied. First, the spatial distribution of opinions after 100 steps of simulation was analyzed. The simulations showed how clusters of opinion are formed depending on the flow of information in the agents’ network, as well as after considering the randomness  in forming the opinion. Consensus (one large cluster) was possible for $K=2$ (two opinions) for low values $\alpha$ ($\alpha=1$, 2), when agents formed their opinions also contacting more distant agents (that is, when the flow of information was good in the whole  community). 

Generally, the greater the $\alpha$  and $T$, the more clusters, i.e. groups of agents with the same opinions (less polarization of opinions in the groups). Of course, it should be noted that the appearance of noise in the system ($T = 1$) slightly orders the system in relation to the situation when $T=0$ (lack of randomness). This is in accordance with other studies \citep{PhysRevE.75.045101,Biondo2013,Shirado2017,2002.05451}, in which a low noise level has brought order to the system.
With the formation of opinion clusters, the probability of sustaining opinion is closely related. This probability is greater within the clusters than at their borders and it is larger in the larger clusters (\Cref{fig:pso}). This leads to the disappearance of small clusters, and thus to reduction of the number of clusters.

The formulation of opinions  is also described by the number and size of clusters of opinion. The clustering of opinions is influenced by both the level of randomness in agents’ decisions and the influence
coming from neighbors. Generally, the size of the largest cluster of opinions decreases with the increase of $\alpha$. This, of course, corresponds with the spatial distribution of opinions (see \Cref{fig:xi}). Furthermore, the number of clusters increases with $\alpha$, i.e. the smaller the influence of all agents in the network on the selected agent, the more difficult it is for forming clusters of opinions. An interesting phenomenon occurs in the case of analysis of the impact of $T$ (for $T > 0$). The probability of a new cluster forming (with different opinion) inside another cluster is then small and the probability of changing opinions is greater at the border between two clusters than inside them. Additionally, the probability of changing the agent’s opinion in a smaller cluster is greater than in the larger one, which leads to the disappearance of small clusters, or in general to a reduction in the number of clusters. 
In summary, the simulations showed that opinion formation and spread is influenced by both flow of information between agents (interactions outside the closest neighbors) and randomness (irrationality) in adopting opinions (what is shown in \Cref{tab:summary}). 

\begin{table*}
\caption{\label{tab:summary}The impact of both $T$ and $\alpha$ on the formation and spread of opinions.}
 	\begin{tabular}{p{5cm}p{4.5cm}p{4.5cm}}
	\hline\hline
Impact on &	for $T>0$&	$\alpha$\\ \hline
Spatial distribution of opinions &	The greater the $T$, there are more and more clusters&	The greater the $\alpha$, there are more and more clusters\\ \hline
Probability of sustaining opinion&	The higher $T$, the less probability
of sustaining the opinion& 	The higher $\alpha$, the chances of sustaining the current opinion of agents decrease\\ \hline
Clustering of opinions&	The number of clusters increases with $T$&
	The number of clusters increases with $\alpha$.
The size of the largest cluster decreases with the increase of $\alpha$\\ \hline
Summary&	Generally randomness (irrationality) hinders polarization of opinions in groups&	The better the flow of information in the community, the easier it is to form and spread opinions, which can also lead to consensus\\ \hline\hline
\end{tabular}
\end{table*}

Better information flow, i.e. contacts with more agents in the network representing the community facilitates the formation of opinion and its spread. In the case of small values of $\alpha$ (when information flow is very good), the result of the simulation is a consensus, as in most sociophysical models of social dynamics \citep{RevModPhys.81.591}. For large values of $\alpha$ (when opinions are consulted only in the close neighborhood), the polarization of opinions is weak and there are many small groups of agents with the same opinion. In addition, the presence of many small clusters causes a lower probability of sustaining opinion for individual agents, i.e. they change their opinions more often. 

 As mentioned earlier, the better the flow of information in the community, the better the opinion spreads and the more often there is consensus and polarization of opinions. Also, \cite{Bjorkman2004} and \cite{Tsai2002} emphasize that good communication and wide interaction channels favor the exchange of knowledge and information. In organization, good spread of opinions and information is particularly important in the case of organizational change, organizational culture and knowledge transfer. Managers should therefore ensure good information flow between members of the organization by creating both formal information flow channels and informal communication networks. Moreover, they should make sure that the number of members in the organization who act in a random and unpredictable manner is small, because the more randomness in the organizational behavior, the less chance of polarization of opinions and consensus appearing.




\end{document}